\documentclass[11pt]{article}

\usepackage[utf8]{inputenc}
\usepackage[T1]{fontenc}
\usepackage{hyperref}
\usepackage{url}
\usepackage{booktabs}
\usepackage{graphicx}
\usepackage{amsmath}
\usepackage[margin=1in]{geometry}
\usepackage{enumitem}
\usepackage{authblk}
\usepackage{xcolor}

\definecolor{manifestocolor}{RGB}{17,116,230}

\title{\textbf{The AI-Native Large-Scale Agile Software Development Manifesto}}

\author[1,2]{Ricardo Britto\thanks{ricardo.britto@ericsson.com}}
\author[1]{Fredrik Palmgren\thanks{fredrik.palmgren@ericsson.com}}
\author[1]{Nishrith Saini\thanks{nishrith.saini@ericsson.com}}
\author[1]{Marcus Ohlin\thanks{marcus.ohlin@ericsson.com}}

\affil[1]{Ericsson, Sweden}
\affil[2]{Blekinge Institute of Technology, Sweden}

\date{}

\begin{document}

\maketitle

\begin{abstract}
Despite the widespread adoption of agile methods, achieving true agility at scale remains elusive. Large-scale agile frameworks remain largely human-centric and manual, relying on coordination meetings, artifact synchronization, and role-based handoffs that inhibit real-time adaptation. Meanwhile, rapid advances in AI, particularly large language models, have begun transforming software engineering, yet their potential for organizational-level agility remains underexplored. We present the AI-Native Large-Scale Agile Software Development Manifesto: a set of values and principles that redefine how large-scale software development is organized when AI becomes a first-class participant rather than a peripheral tool. The manifesto is grounded in six principles, parallel processes, intent-driven teams, living knowledge, verification-first assurance, orchestrated agent workforces, and reusable blueprints, that together shift development from a meeting-driven, document-heavy, sequential process to an intelligent, adaptive, continuously learning system.
\end{abstract}

\section*{The Manifesto}

\noindent We are uncovering better ways of developing software at scale by doing it and helping others do it. Through this work, we have come to value:

\vspace{1em}

\begin{center}
\large
\textbf{Shared context and living knowledge} over meetings and static documents\\[0.8em]
\textbf{Human intent and AI execution} over manual handoffs across roles\\[0.8em]
\textbf{Parallel, continuous verification} over sequential phase gates\\[0.8em]
\textbf{Orchestrated AI workforces} over isolated tool assistance\\[0.8em]
\textbf{Reusable blueprints with local adaptation} over rigid frameworks or reinvention\\[0.8em]
\end{center}

\vspace{1em}

\noindent That is, while there is value in the items on the right, we value the items on the left more.

\section{Introduction}

Despite the widespread adoption of agile methods, organizations continue to struggle to achieve ``true agility'' at scale, where velocity, responsiveness, alignment, and quality are sustained across multiple teams, systems, and domains. Evidence indicates critical impediments, including fragmented coordination, knowledge silos, inconsistent quality assurance, and difficulty maintaining shared understanding across distributed teams~\cite{dingsoyr2014, rolland2016}. As software-intensive systems grow in complexity, these issues amplify, leading to slower feedback cycles, costly integration challenges, and the erosion of system-level agility.

Large-scale agile frameworks (e.g., SAFe~\cite{safe2020} and LeSS~\cite{larman2016}) remain largely human-centric and manual, relying heavily on coordination meetings, artifact synchronization, and role-based handoffs, which inhibit real-time adaptation and the necessary velocity and responsiveness to cater to the needs of the current fast-paced world~\cite{ambler2012, larman2016, safe2020}.

In parallel, rapid advances in Artificial Intelligence (AI), particularly the emergence of large language models (LLMs), have begun transforming the way software is engineered. AI-assisted coding tools such as GitHub Copilot, Amazon Kiro, and Claude Code have demonstrated substantial productivity gains at the individual developer level~\cite{guo2023}. Yet their potential for transforming organizational-level agility in large-scale contexts remains underexplored.

This manifesto and its accompanying principles outline the vision for \textbf{AI-Native Large-Scale Agile Software Development}, a fundamental rethink that aims to realize true agility at scale within a cohesive, AI-augmented development ecosystem. Our vision can be summarized as:

\begin{quote}
\textit{Human intent, machine velocity, where human experts direct and AI agents deliver.}
\end{quote}

The AI-native approach transforms large-scale software development into an intelligent, adaptive, and learning system characterized by:
\begin{itemize}[nosep]
    \item Humans set the intents, and a context-aware, knowledge-rich coordination approach handles interactions with an AI-agentic workforce, eliminating unnecessary meetings.
    \item Parallel development with systemic consistency, where AI agents co-create, verify, and evolve software.
    \item Transparent and explainable automation, grounded in a semantic layer.
    \item Reusable blueprints that enable systematic introduction and evolution of the AI-native approach.
\end{itemize}

\begin{figure}[ht]
    \centering
    \includegraphics[width=\textwidth]{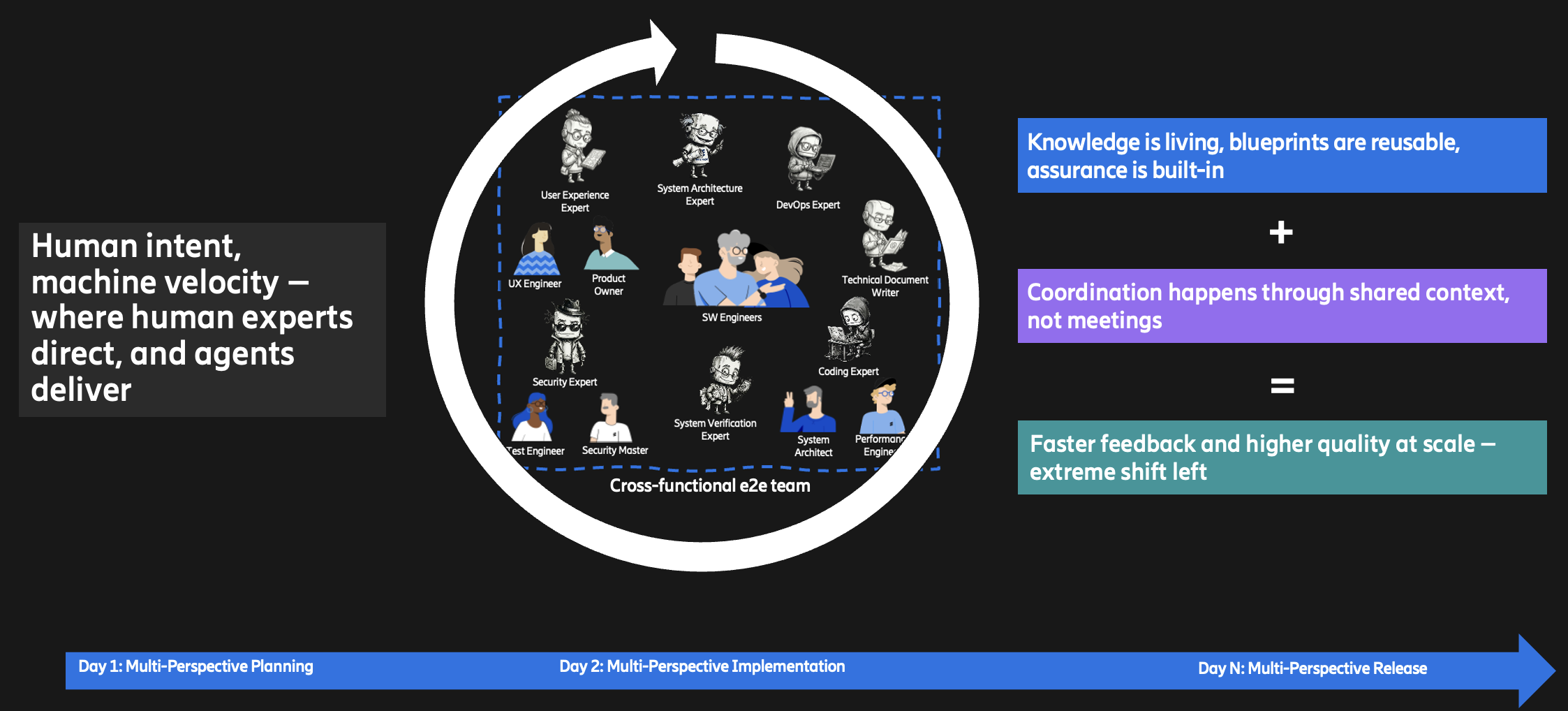}
    \caption{AI-native R\&D overview: human intent and machine velocity in cross-functional end-to-end teams, where knowledge is living, blueprints are reusable, assurance is built-in, and coordination happens through shared context, resulting in faster feedback and higher quality at scale.}
    \label{fig:overview}
\end{figure}

This rethink manifests across three reinforcing dimensions. First, the fundamental development activities, analysis, architecture, design, coding, testing, release, and deployment, remain the same, but the paths through them shift from sequential handoffs to a continuous, intent-driven spiral with fewer gates and faster feedback loops. Second, artifacts evolve from static documents into living, structured knowledge that serves both humans and AI agents throughout the life-cycle. Third, teams transform into human--AI collaborations where agents progressively move from supporting roles to autonomous contributors, while humans move up in abstraction, from execution to direction. Figure~\ref{fig:overview} illustrates this vision.

\section{Principles Behind the Manifesto}

Our manifesto is grounded in six principles that form a system rather than a series of independent improvements (see Table \ref{tab:principles}). Together, they represent a shift from development as a meeting-driven, document-heavy, sequential process to development as an intelligent, adaptive, continuously learning system.

\begin{table}[!ht]
\centering
\caption{AI-native large-scale agile software development principles.}
\label{tab:principles}
\begin{tabular}{@{}p{3.5cm}p{10cm}@{}}
\toprule
\textbf{Principle} & \textbf{Description} \\
\midrule
The New Process & Replace linear handoffs with parallel work. Humans and AI plan, build, and verify simultaneously. Coordination through shared context, not meetings. \\
The New Teams & Humans direct, AI executes. Engineers define intent and verify output rather than writing all the code. Human in Control, not Human in the Loop. \\
The New Artifacts & Static docs are replaced by AI-ergonomic, version-controlled specs that agents can read, act on, and update. A unified semantic layer keeps knowledge up to date. \\
The New Assurance & Tests are generated from specs before code is written. BDD scenarios define ``done'' upfront. AI agents manage scenario explosion at scale. \\
The Extended Workforce & AI agents, as a coordinated team of specialists, work closely together with human engineers. \\
The New Reuse & Teams publish reusable blueprints (personas, specs, skills, knowledge graphs). Other teams adopt and adapt them locally. Reuse without rigidity. \\
\bottomrule
\end{tabular}
\end{table}

In the rest of this section, we provide more details about the six principles underlying our AI-native vision.

\subsection{The New Process: Parallel by Default}

The most visible change is the end of sequential work. In an AI-native organization, planning, implementation, and verification happen concurrently, not as phases that follow each other, but as activities that run in parallel, informed by the same shared context.

This is possible because the cost of each activity drops dramatically. When AI agents can draft code, generate tests, and analyze specs in minutes rather than days, the economics of the process flip. You no longer need to carefully gate every step to protect the cost of human effort. Instead, you iterate fast, correct quickly, and let the workflow adapt.

The few gates that remain must be automatable, agent-assessable quality checks rather than manual approval meetings. The goal is to shift left, reduce silos, and minimize the number of investment decisions that slow things down. Coordination happens through shared context, not through calendars.

\subsection{The New Teams: Intent \& Oversight}

If the process changes this fundamentally, so must the teams that operate within it.

The traditional developer writes code. The AI-native developer defines intent. The shift is from execution to direction, from being the hands on the keyboard to being the mind that decides what should be built, provides the context for how it should work, and verifies that the result is correct.

Engineers become designers of environments, guardrails, and controls. They shape the conditions under which AI agents operate, rather than doing the work those agents now handle. This is not a reduction in skill; it is a different and arguably harder skill. Knowing what to build and recognizing whether the output is right requires deep domain expertise.

The governance model also shifts. \textit{Human in the Loop}, where every step requires human approval, becomes \textit{Human in Control}. The difference is subtle but critical: you retain full authority over the outcome without becoming a bottleneck in the process. Corrections after the fact are cheap. Waiting for approval at every step is expensive.

\subsection{The New Artifacts: Living Knowledge}

Every organization runs on knowledge, requirements, architecture decisions, domain rules, API contracts, and process definitions. Today, most of that knowledge lives in static documents: Word files, PDFs, wiki pages, slide decks. These artifacts have a half-life measured in weeks. The moment they are written, they begin to decay.

AI-native development replaces static documents with living knowledge. Specifications become the primary source of truth, version-controlled, traceable, and replaceable. You can see exactly which spec produced which output and replay the entire chain when something needs to change.

All artifacts are restructured to be AI-ergonomic, designed for both human understanding and machine consumption. Knowledge is fused into a unified semantic layer aligned with a shared ontology, enabling agents to autonomously find, interpret, act on, and update organizational knowledge.

This creates a feedback loop that static documents never could: agents consume the knowledge to do their work, and their work updates the knowledge. The artifacts stay alive because the system that uses them also maintains them.

\subsection{The New Assurance: Verification-First}

In traditional development, testing comes after implementation. You build something, then check if it works. The later you find a problem, the more expensive it is to fix.

Verification-first inverts this. Tests are generated from specifications before any code is written. Acceptance criteria, black-box tests, and unit tests all validate the spec, not just the implementation. Requirements are locked in upfront, and the tests become a contract that the code must satisfy.

This naturally extends to Behavior-Driven Development (BDD), where domain behavior is embedded in testable specifications using a language that business stakeholders, developers, and testers all share. The Given--When--Then format ensures that everyone agrees on what ``correct'' means before a single line of code is generated.

The challenge with BDD has always been scale \cite{IRSHAD2021}. Across many teams and features, scenarios multiply, overlap, and contradict each other. In an AI-native setting, agents take over this management, reasoning about scenario overlap, detecting inconsistencies, and maintaining scenario hierarchies across the entire system. What was previously a manual coordination burden becomes an automated, continuously maintained assurance layer.

\subsection{The Extended Workforce: Orchestrated Agents}

Behind all of this is an extended workforce working closely together with human engineers: AI agents organized not as a single all-knowing assistant but as a coordinated team of specialists.

A supervisor agent orchestrates role-based subagents, each with a clearly defined scope, just as a human manager delegates to team members with specific expertise. One agent handles requirements analysis, another generates tests, another reviews code, and another manages knowledge extraction. Each operates within its own bounded context, focused on what it does best.

This structure solves a fundamental limitation of AI: context overflow. A single agent trying to handle everything eventually loses track, and quality degrades. By splitting work across scoped subagents, with session resets, delegation protocols, and externalized context, the system stays sharp regardless of how large or complex the project becomes.

\begin{figure}[ht]
    \centering
    \includegraphics[width=0.85\textwidth]{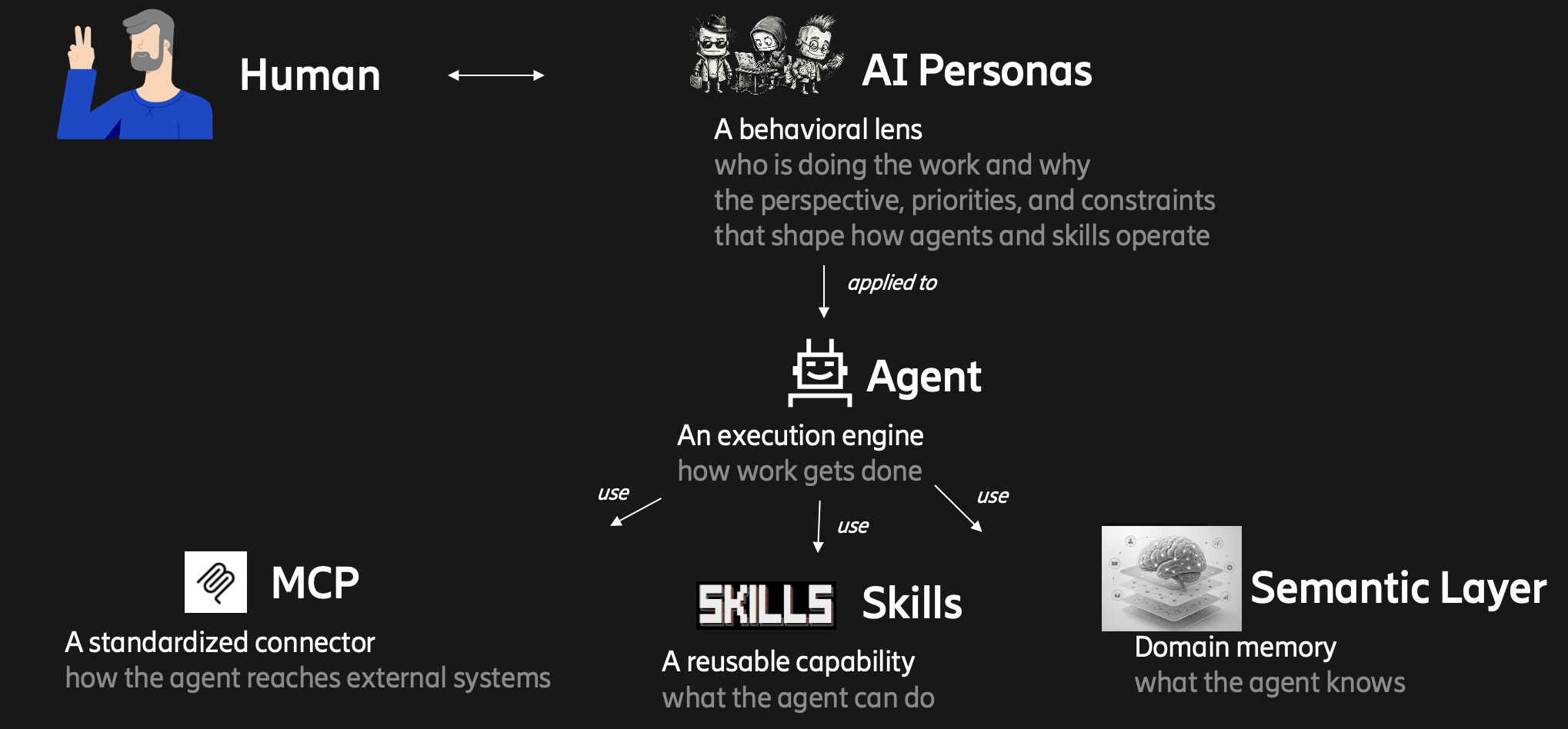}
    \caption{The relationship between humans, AI personas, agents, skills, MCP connectors, and the semantic layer. Personas define the behavioral lens (who and why), agents are the execution engines (how), and skills provide reusable capabilities (what). Agents connect to external systems via MCP and draw on the semantic layer as domain memory.}
    \label{fig:persona-agent-skill}
\end{figure}

Three constructs make this operational (Figure~\ref{fig:persona-agent-skill}):
\begin{itemize}[nosep]
    \item A \textbf{Persona} is a role identity, a set of competences that defines who is doing the work and why they make certain decisions, including the perspective, judgment, priorities, and constraints that shape how work is approached.
    \item An \textbf{Agent} is a digital worker, the execution engine that carries out tasks within a defined scope. It has defined instructions, a set of Skills, is executed in an environment, and utilizes an AI model.
    \item A \textbf{Skill}~\cite{agentskills} is a reusable capability, a narrow, well-defined function that an agent can perform.
\end{itemize}

\subsection{The New Reuse: Blueprints \& Local Adaptation}

The final principle addresses how all this scales across an organization.

Rather than having every team develop AI-native practices from scratch, the organization publishes reusable blueprints on a centralized platform. These blueprints encode proven patterns: how to structure a requirements review agent, set up verification workflows, and organize knowledge for a particular domain.

Nevertheless, blueprints are starting points, not mandates. Teams adopt them and adapt them locally, tailoring them to their specific domain, technology stack, and workflows. This balance between organizational consistency and team autonomy is essential. A one-size-fits-all approach ignores the reality that teams face different challenges. Pure local invention ignores the reality that most challenges have already been solved somewhere else.

The result is an organization that learns collectively while acting locally, where every team benefits from others' experience without being constrained by it.

\section{The AI-Native Blueprint}

The AI-native approach represents a paradigm shift in achieving agility at scale. Rather than treating AI as a coding assistant or an isolated productivity tool, it embeds AI within the socio-technical fabric of software development, acting as context-aware personas that collaborate with human experts across the life-cycle. At its core, it combines behavior-driven development (BDD), parallel agile, and multi-agent collaboration, underpinned by a semantic layer that enables AI personas to reason about and act on the evolving software system.

\begin{figure}[ht]
    \centering
    \includegraphics[width=\textwidth]{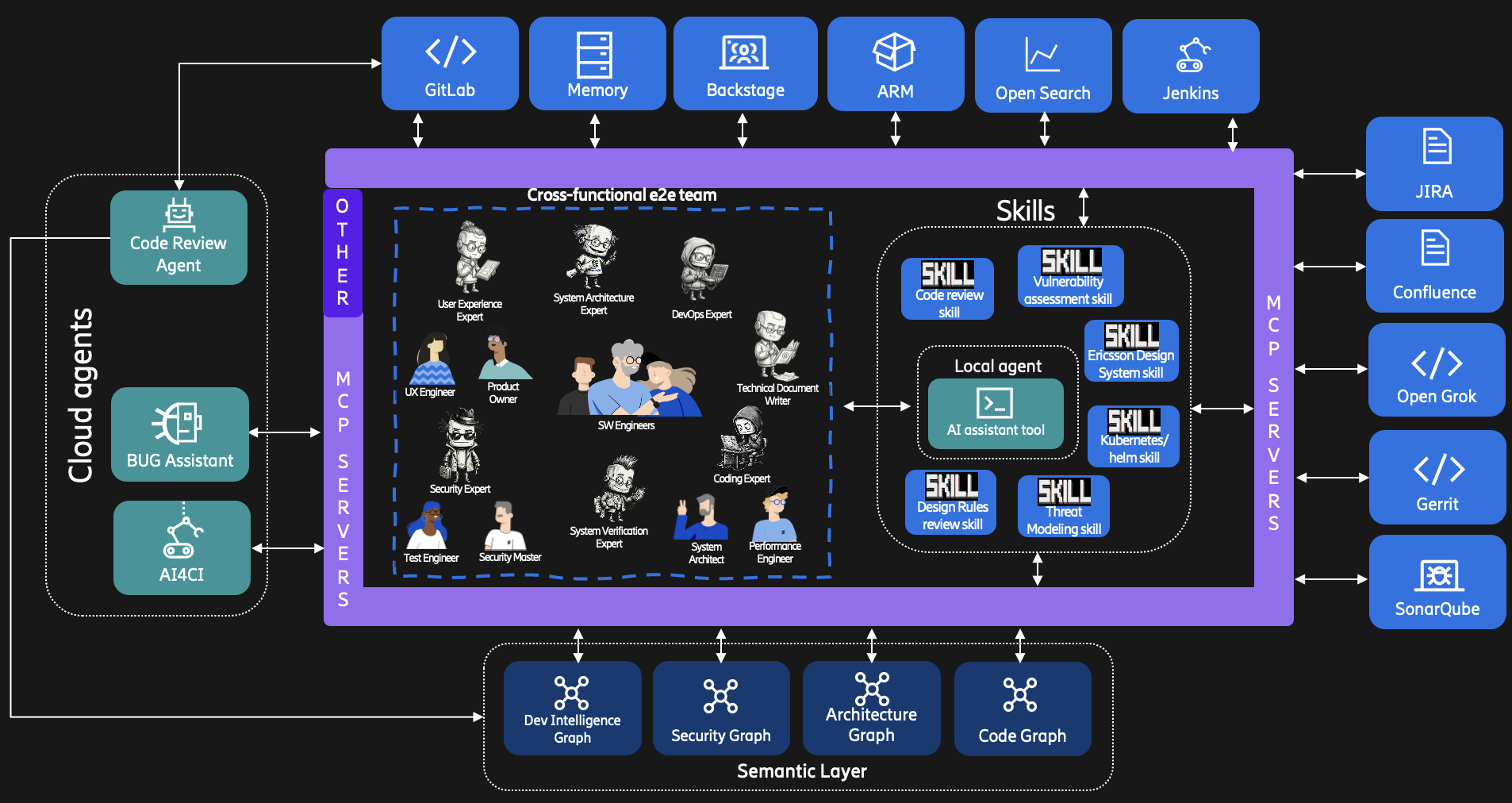}
    \caption{AI-native R\&D blueprint: a cross-functional end-to-end team supported by a local agent with reusable skills, cloud agents, MCP servers connecting to development tools, and a semantic layer comprising domain-specific knowledge graphs.}
    \label{fig:blueprint}
\end{figure}

The approach replaces sequential handoffs among specialized roles with contextually collaborative and concurrent processes, where multiple perspectives, security, architecture, verification, UX, and operations engage simultaneously at each stage of the life-cycle. The blueprint (Figure~\ref{fig:blueprint}) describes how roles, processes, and artifacts interconnect within the AI-native ecosystem.

Each development cycle begins with requirement analysis and solution specification, which feed into concurrent workstreams involving humans and AI agents. Unlike traditional phase-based flows, these roles collaborate in overlapping cycles. Example personas include:

\begin{itemize}[nosep]
    \item \textbf{System Architecture Expert} ,  contributes system baseline models and maintains architectural integrity.
    \item \textbf{Security Expert},  dynamically runs vulnerability analyses, penetration tests, and risk reviews.
    \item \textbf{Verification Expert},  ensures that requirements and solutions are consistently testable and verified early.
    \item \textbf{UX Expert} provides early prototypes and user-behavior validation from the user perspective.
    \item \textbf{DevOps Expert},  assures deployability and serviceability from the start.
\end{itemize}

AI agents continuously use a shared semantic layer, ensuring that both humans and agents work with a consistent, up-to-date system context. This tightly coupled but contextually distributed setup mirrors the Parallel Agile principle of enabling concurrent workstreams with synchronized validation, supported by behavior-driven specifications.

The composition of human--AI teams is expected to evolve over time. Initially, agents support human engineers in an assistive capacity. As shared knowledge and structured artifacts mature, agents begin collaborating more broadly, with multiple agents with different perspectives working together as coordinated teams. As this evolution progresses, humans move up an abstraction level, from writing code to specifying intent, from managing individual tasks to shaping strategy, and from reviewing implementation details to governing outcomes.

\subsection{The Semantic Layer}

The semantic layer forms the cognitive foundation of the AI-native approach. It integrates multiple domain-specific knowledge graphs (KGs) to provide a unified contextual representation of the system:

\begin{itemize}[nosep]
    \item \textbf{Architecture KG}: Captures product architecture, dependencies, and interfaces.
    \item \textbf{Code KG}: Synthesizes code elements, relationships, and abstractions into a semantic network.
    \item \textbf{Security KG}: Represents vulnerabilities, threat models, and security dependencies.
    \item \textbf{Dev-Intelligence KG}: Connects process metrics, test results, and tool metadata for engineering analytics.
\end{itemize}

Together, these graphs form a shared memory for both humans and AI personas, allowing contextual reasoning, traceability, and explainable decision support. Through the semantic layer, AI personas can perform:

\begin{itemize}[nosep]
    \item \textbf{Cross-domain impact analysis}, by traversing linked entities (e.g., requirement $\rightarrow$ code $\rightarrow$ test $\rightarrow$ defect).
    \item \textbf{Fact-grounded reasoning}, reducing hallucination risk common in LLMs.
    \item \textbf{Knowledge-driven automation}, enabling the ``analyze $\rightarrow$ decide $\rightarrow$ act'' workflow loops characteristic of agentic systems.
\end{itemize}

This enables automated cross-perspective reasoning, for example: detecting when a new requirement impacts existing architectural constraints, suggesting test case updates when the code or behavior specification evolves, and highlighting security vulnerabilities based on architectural dependencies.

\section{Platform Engineering as an Enabler}

Platform Engineering is an essential element for realizing the AI-native vision. It has traditionally focused on creating seamless developer experiences by providing standardized access to tools, environments, and assets. As AI agents become integral to software development workflows, Platform Engineering's role expands to serve both human engineers and AI agents.

Just as engineers need intuitive interfaces and clear pathways to infrastructure, AI agents require structured access to the same resources to perform their tasks effectively. This dual responsibility positions Platform Engineering as the critical bridge between human intent and AI execution, ensuring both audiences can interact with systems efficiently and securely.

The concept of ``golden paths'', automated implementations of recommended practices, becomes even more valuable in this AI-augmented landscape. These golden paths, typically built from proven blueprints, serve as guardrails that guide both engineers and agents toward best practices while reducing cognitive load and decision fatigue. For human engineers, golden paths simplify complex workflows into repeatable patterns. For AI agents, they provide well-defined interfaces and constraints that enable autonomous operation within safe boundaries.

\section{Future Directions}

The vision outlined here is expected to evolve as technology advances, especially given the current pace of AI development. It is essential to revisit the manifesto regularly and ensure it accommodates relevant advancements. Of special interest is the emerging trend of pushing abstraction further, using ``dark software factories'' where AI agents are the only ones coding, while humans specify behaviors and intents, along with the necessary testing artifacts.

As the field matures, we anticipate:
\begin{itemize}[nosep]
    \item Increasingly autonomous agent teams that require less human oversight for routine development tasks.
    \item Richer semantic layers that capture not just system structure but organizational learning and decision rationale.
    \item Cross-organizational blueprint ecosystems where industry-wide patterns accelerate adoption.
    \item New professional roles focused on agent orchestration, knowledge engineering, and AI governance.
\end{itemize}

\section{Conclusion}

The AI-Native Large-Scale Agile Software Development Manifesto represents a call to rethink how we organize software development at scale. The core insight is simple: when AI can perform development activities at machine speed, the bottleneck shifts from execution to direction. The manifesto's values and principles address this shift by placing human intent at the center while leveraging AI agents as a coordinated, context-aware workforce.

This is not about replacing humans with machines. It is about elevating human work to where it creates the most value, defining what should be built, why it matters, and whether the result is correct, while AI handles the how at a pace and scale that humans alone cannot match.

\bibliographystyle{IEEEtran}
\bibliography{references}

\begin{thebibliography}{1}
\providecommand{\url}[1]{#1}
\csname url@samestyle\endcsname
\providecommand{\newblock}{\relax}
\providecommand{\bibinfo}[2]{#2}
\providecommand{\BIBentrySTDinterwordspacing}{\spaceskip=0pt\relax}
\providecommand{\BIBentryALTinterwordstretchfactor}{4}
\providecommand{\BIBentryALTinterwordspacing}{\spaceskip=\fontdimen2\font plus
\BIBentryALTinterwordstretchfactor\fontdimen3\font minus \fontdimen4\font\relax}
\providecommand{\BIBforeignlanguage}[2]{{%
\expandafter\ifx\csname l@#1\endcsname\relax
\typeout{** WARNING: IEEEtran.bst: No hyphenation pattern has been}%
\typeout{** loaded for the language `#1'. Using the pattern for}%
\typeout{** the default language instead.}%
\else
\language=\csname l@#1\endcsname
\fi
#2}}
\providecommand{\BIBdecl}{\relax}
\BIBdecl

\bibitem{dingsoyr2014}
T.~Dingsøyr, N.~B. Moe, T.~E. Fægri, and E.~A. Seim, ``Exploring software development at the very large scale: a revelatory case study and research agenda for agile method adaptation,'' \emph{Empirical Software Engineering}, vol.~23, no.~1, pp. 490--520, 2018.

\bibitem{rolland2016}
C.~Rolland, N.~Prakash, and C.~Ben~Achour, ``Challenges of large-scale agile: a systems perspective,'' \emph{Journal of Systems and Software}, vol. 125, pp. 1--14, 2016.

\bibitem{safe2020}
{Scaled Agile Inc.}, ``{SAFe} 5.0 framework overview,'' \url{https://scaledagileframework.com}, 2020.

\bibitem{larman2016}
C.~Larman and B.~Vodde, \emph{Large-Scale Scrum: More with LeSS}.\hskip 1em plus 0.5em minus 0.4em\relax Addison-Wesley, 2016.

\bibitem{ambler2012}
S.~W. Ambler and M.~Lines, \emph{Disciplined Agile Delivery: A Practitioner's Guide to Agile Software Delivery in the Enterprise}.\hskip 1em plus 0.5em minus 0.4em\relax IBM Press, 2012.

\bibitem{guo2023}
P.~J. Guo, ``The state of {AI}-assisted programming: emerging opportunities and challenges,'' \emph{Communications of the ACM}, vol.~66, no.~11, pp. 68--77, 2023.

\bibitem{IRSHAD2021}
M.~Irshad, R.~Britto, and K.~Petersen, ``Adapting behavior driven development (bdd) for large-scale software systems,'' \emph{Journal of Systems and Software}, vol. 177, p. 110944, 2021.

\bibitem{agentskills}
{AgentSkills}, ``Agent skills,'' \url{https://agentskills.io/}.

\end{thebibliography}

\end{document}